\documentclass[journal=apchd5,manuscript=article]{achemso}

\usepackage[version=3]{mhchem} 

\author{Marion Castilla}
\author{Silvère Schuermans}
\author{Davy Gérard}
\author{Jérôme Martin}
\author{Thomas Maurer}
\affiliation[L2n]
{Light, nanomaterials, nanotechnologies (L2n), Université de Technologie de Troyes, CNRS EMR 7004, 12 rue Marie Curie, Troyes 10000, France}
\author{Uri Hananel}
\author{Gil Markovich}
\affiliation{School of Chemistry, Tel Aviv University, Tel Aviv 6997801, Israel}
\author{Jérôme Plain}
\email{jerome.plain@utt.fr}
\author{Julien Proust}
\email{julien.proust@utt.fr}
\affiliation[L2n]
{Light, nanomaterials, nanotechnologies (L2n), Université de Technologie de Troyes, CNRS EMR 7004, 12 rue Marie Curie, Troyes 10000, France}

\title{Colloidal synthesis of crystalline aluminum nanoparticles for UV plasmonics}


\begin{document}

\begin{tocentry}

 \includegraphics[width=9cm]{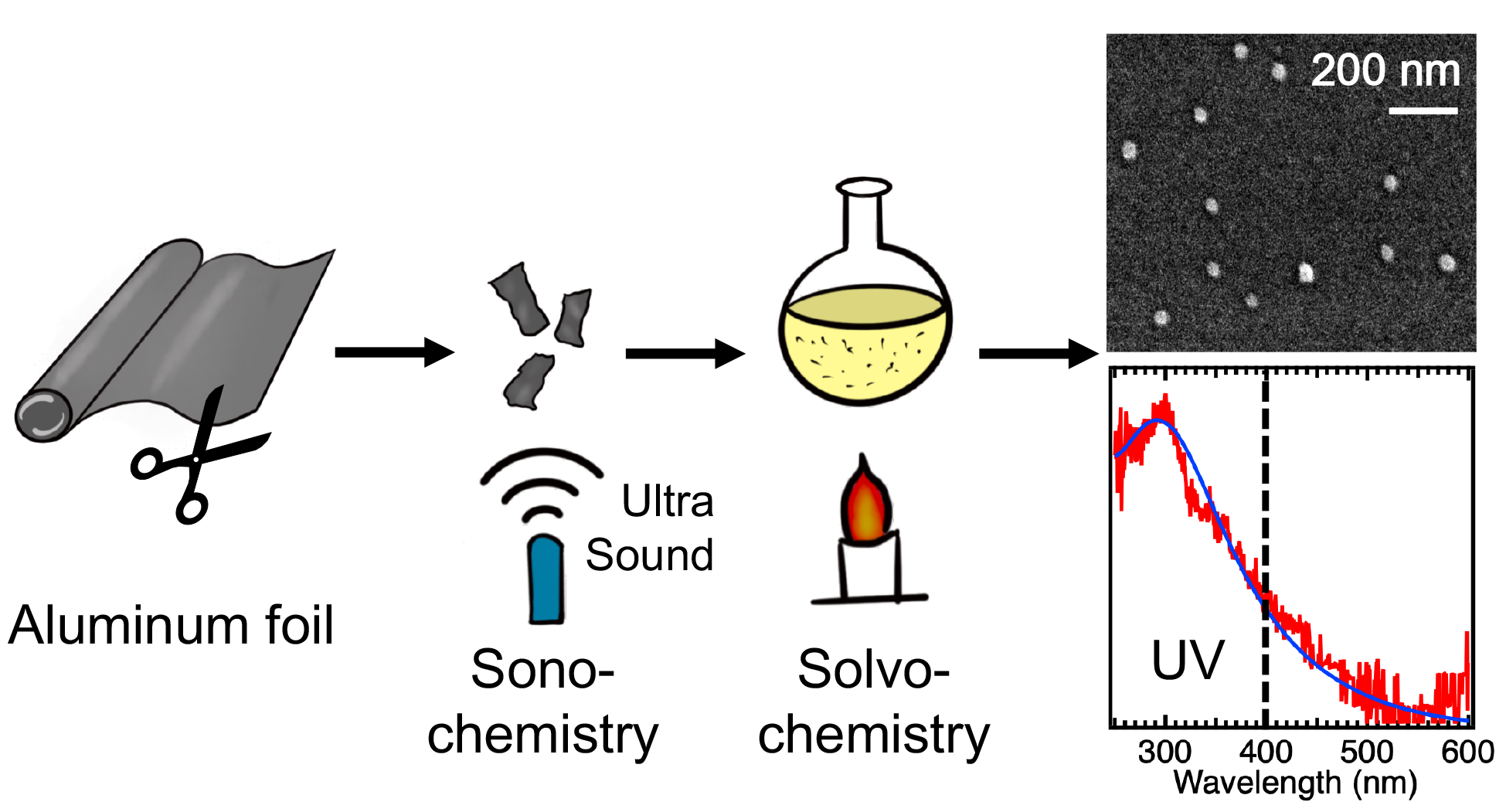}
  \label{TOC}
  
\end{tocentry}

\begin{abstract}
 Numerous applications of nanotechnologies rely on the wide availability of high-quality crystalline nanoparticles (NPs). Although the chemical synthesis of noble metal nanoparticles (gold and silver) is well mastered, pushing the optical response of metallic nanoparticles towards the ultraviolet requires other materials, such as aluminum. Although a few demonstrations of chemical synthesis of Al nanoparticles have been reported so far, an elaboration path allowing for a wide range of NP size that is compatible with mass production has yet to be demonstrated. In this article, we report on the production of spherical, size-controlled crystalline Al NPs starting from commercial Al foils and without the use of a catalyst. The proposed method combines sonochemistry with solvochemistry and is fully up-scalable. The obtained NPs have a 10-100 nm crystalline Al core surrounded by a thin alumina shell, allowing long-term stability in ethanol. Well-defined plasmonic resonances in the UV and visible ranges are experimentally evidenced on single nanoparticles.
\end{abstract}

\section{Introduction}
Aluminum is emerging as an appealing plasmonic material \cite{knight2014aluminum, gerard2014al}, notably due to its capability to sustain plasmonic resonances in the ultraviolet (UV) \cite{mcmahon2013plasmonics,maidecchi2013} -- a spectral range where noble metals behave as absorbing materials. Moreover, aluminum is an abundant, non-critical \cite{graedel2015}, and CMOS-compatible \cite{olson2014} material. To obtain Al nanostructures, top-down fabrications methods are currently the most common approaches \cite{martin2014fabrication}. However, for mass scale fabrication, chemical syntheses are highly desirable. Furthermore, chemical approaches generally yields crystalline and shape-controlled nanoparticles, giving better quality (i.e., higher quality factor) plasmonic resonances. While the chemical synthesis of gold and silver nanoparticles is common knowledge, the synthesis of Al nanoparticles is not so easy, with only a handful of reports published so far \cite{mcclain2015, clark2019, renard2020uv, lu2021}. The underlying reason is the fact that the method used for noble metals, the reduction of a metallic salt, cannot be directly translated to aluminum because the classical reducers cannot reduce aluminum salt. Hence, other approaches are required. Haber and coworkers \cite{haber1998} demonstrated the first chemical synthesis of Al nanocrystals in 1998 by a reaction between lithium aluminium hydride and aluminum chloride. In 2009, Paskevicius \textit{et al.} \cite{paskevicius2009mechanochemical} adapted the synthesis for smaller nanoparticles for deuterium detection. However, the resulting particles were in an organic matrix and difficult to purify. A huge leap forward was made by McClain and coworkers \cite{mcclain2015} in 2015, who demonstrated the synthesis of Al nanoparticules in tetrahydrofuran using the chemical reduction of aluminium hydride by cumyl dithiobenzoate-terminated polystyrene as ligand. The size of the obtained Al particles was sub-micronic and therefore too large for UV-plasmonic applications. Smaller nanoparticles were obtained by Renard and coworkers \cite{renard2020uv} in 2020 by a modification of this previous synthesis adding a silica shell around the aluminum crystals. 

The synthesis of crystalline, spherical Al nanoparticles with a controlled diameter allowing resonances covering the visible and near-UV remains challenging. To be industry-compatible, such an elaboration path should also be up-scalable to mass production and use as a starting point a cheap and widely available material. In 2014, we patented such a path, obtaining Al nanoparticles from the ultra-sonication of commercial aluminum foils \cite{brevetalu}. However, the presence of byproducts in the solution was hindering the usefulness of the method. Note that very recently, a related approach was reported by Lu \textit{et al.} \cite{lu2021}, who demonstrated the creation of crystalline 2D nanocrystals (flakes) after ultrasonic exfoliation of commercial Al foils. The method remains limited to the synthesis of 2D nanoparticles and their size is not under control.   

In this paper, we propose a two-step synthesis of crystalline aluminum nanoparticles without catalyst. Starting from commercial aluminum foils, we first create Al precursors using sonochemistry. We point out that commercial aluminum foil are pure aluminum and without any coating. The aluminum foils are exposed for several hours to high-power ultrasound, generating cavitation bubbles which trigger a sono-chemical reaction leading to the creation of \ce{Al^3+} ions. These ions are then used as precursors in the second step of solvochemistry. After purification of the solution from the first step to remove the largest chunks of the remaining aluminum foil, ethylene glycol is added. The resulting mixture is then annealed, creating Al clusters that slowly grow with time. The diameter of the Al nanoparticles is hence directly controlled by the solvochemistry reaction time. A final step of purification allows to separate the Al nanoparticles from the other reaction products and to transfer them in ethanol. The resulting Al NPs are then characterized using electron microscopy and single particle optical spectroscopy.

\section{Results and discussion}
We start by presenting the elaboration path and the characterization of the Al nanoparticles. Note that practical details on the synthesis can be found in the accompanying Supplementary Information as a Supplementary Video. This video shows all the steps of the procedure, to help the interested reader to reproduce the synthesis.

\subsection{sonochemistry}

The bottom-up way to synthesize metallic colloidal nanoparticles usually involves the reduction of metallic precursors. In the case of the synthesis of aluminum nanoparticles, the common precursors are based on aluminum hydride compound \cite{ghorbani2014review}. Each precursor allows the fabrication of a limited range of nanoparticles. For example Haber \textit{et al.} \cite{haber1998} created nanoparticles with diameters ranging from 44 nm to 82 nm with an aluminum hydride amine molecule H$_3$AlNMe$_2$Et in mesitylene. In the proposed method, we develop a single-step synthesis of multiple aluminum precursors using sonochemistry starting from commercial aluminum foils. Sonochemistry is an ultrasonic-based chemistry \cite{luche2013synthetic}. This method can be used to produce nanoparticles \cite{gedanken2004using} by choosing the right compound, such as Kan \textit{et al.} who synthesized Au/Pd nanoparticles in ethylene glycol \cite{kan2003ultrasonic}. Here, we use high power ultrasound in order to generate aluminum precursors from aluminum foils in a specific solvent mix. This step is then a top-down process. 900 mg of aluminum foils are cut into small pieces and put into a flask with 30 mL of anhydrous ethylene glycol (99,8$\%$) and 60 mL of absolute ethanol (99$\%$).

\begin{figure}
\includegraphics[width=8cm]{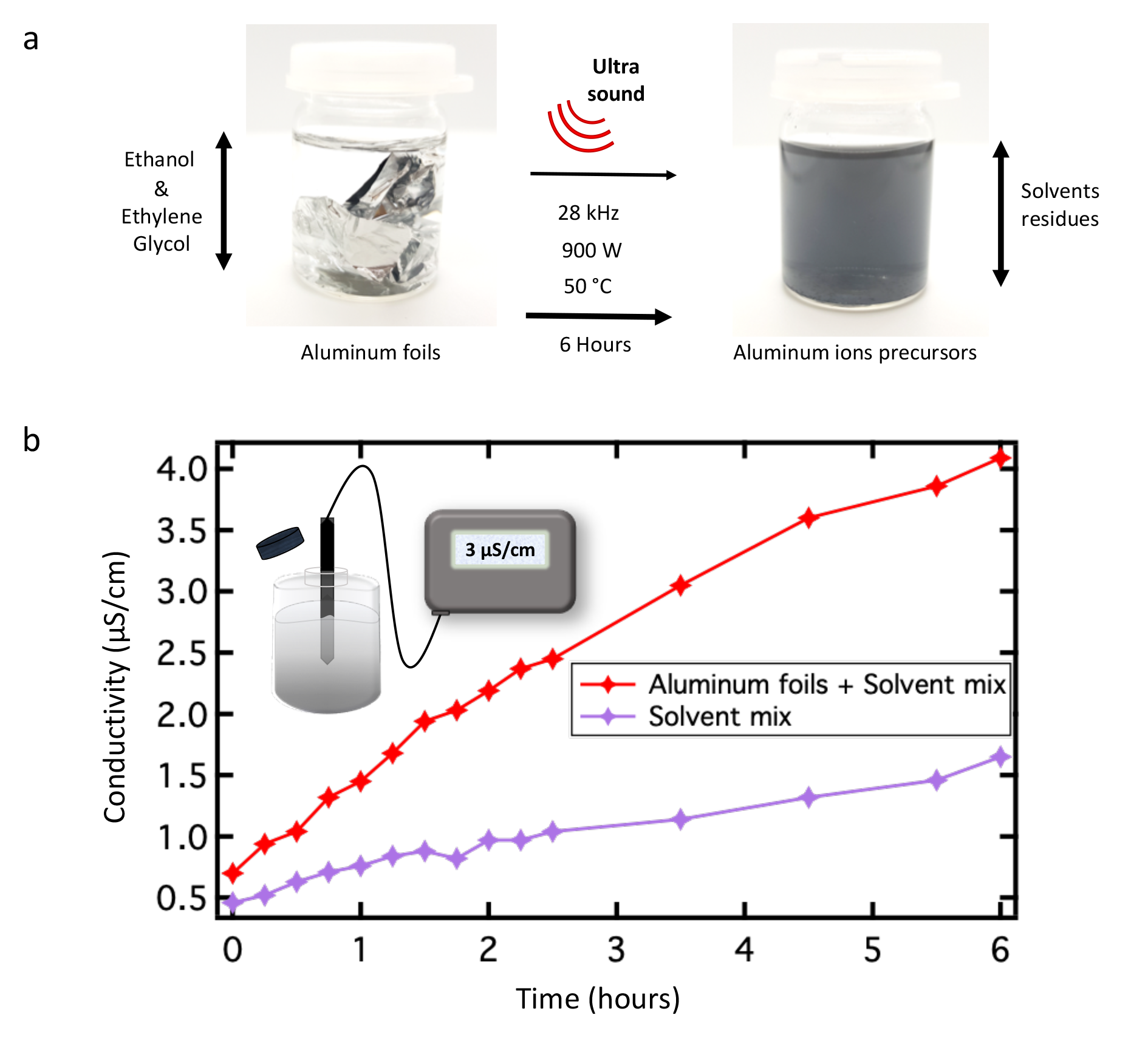}
\caption{(a) Schematic of the sonochemistry process. (b) Conductivity measurements of the solvents mix (with and without aluminum foils) depending on the sonochemistry time.}
\label{fig_conductimetrie}
\end{figure}

The frequency of the ultrasound and the bath temperature are chosen to maximize the creation of cavitation bubbles in the solution \cite{luche2013synthetic}. To this end, we chose a high power ultrasonic generator (Sinaptec, France) at 28 kHz, which was found to be optimal. The previously prepared flask was placed in a 6 L of water at 50 °C  ultrasonic bath for 6 hours, with a power of 900 W (Figure \ref{fig_conductimetrie}a). The initial solution in the flask is transparent with macro aluminum foils. At the end of the ultrasonic process, the macro foils disappear and the solution becomes black, owing to the presence of micro/nano foils and aluminum precursors in the solution. We assume that the majority of the resulting aluminum precursors are single ions \ce{Al^3+}, but the presence of charged clusters cannot be ruled out. The production of aluminum precursors has been investigated by conductivity measurements. In Figure \ref{fig_conductimetrie}b, we plot the conductivity of the reference sample (only the solvent mix) and of the aluminum sample (aluminum foils in the solvent mix) vs the duration of the  ultrasonic process. We observe a higher increase of the conductivity in the case of the sample containing aluminum foils (from 0.7 µS/cm to 4 µS/cm) compared to the case of the bare solvent (from 0.4 µS/cm to 1.7 µS/cm). This increase of the conductivity is due to the creation of Al ions in the solution. To exclude a potential catalyzing effect of the aluminum foils on the solvent, we performed Fourier-transform infrared spectroscopy measurements (see supporting information, Fig. S1). We did not observe any change in the intensity of the peaks associated with solvent degradation. The increase is then due to the creation of aluminum precursors ions during the sonochemistry process. An optimal process time has been found for 6 h with no major increase of the conductivity by increasing this time. After shutting-off the ultrasound, the conductivity of the sample with aluminum continues to increase for 4h (see Fig. S2 in supporting information) due to the modification and stabilization of the precursors. After 10 h, the conductivity is almost stable until 24h. A natural small recombination of the ions in the solution and an agglomeration of the charged clusters at room temperature is observed with a low decrease of the conductivity, down to 3.6 µS/cm after 24 hours. Let us recall that the solvent is a mixture of ethanol and ethylene glycol at 2:1 ratio. The presence of ethanol allows a good cavitation process owing to its low viscosity, while ethylene glycol is essential for the creation and the stabilization of the aluminum precursors in the solution  \cite{gedanken2004using}. After the ultrasonic step, the solvent molecules are turned into various molecules called solvent residues \cite{suslick1983alkane}.

\subsection{Solvochemistry}
\begin{figure}
\centering
\includegraphics[width=8cm]{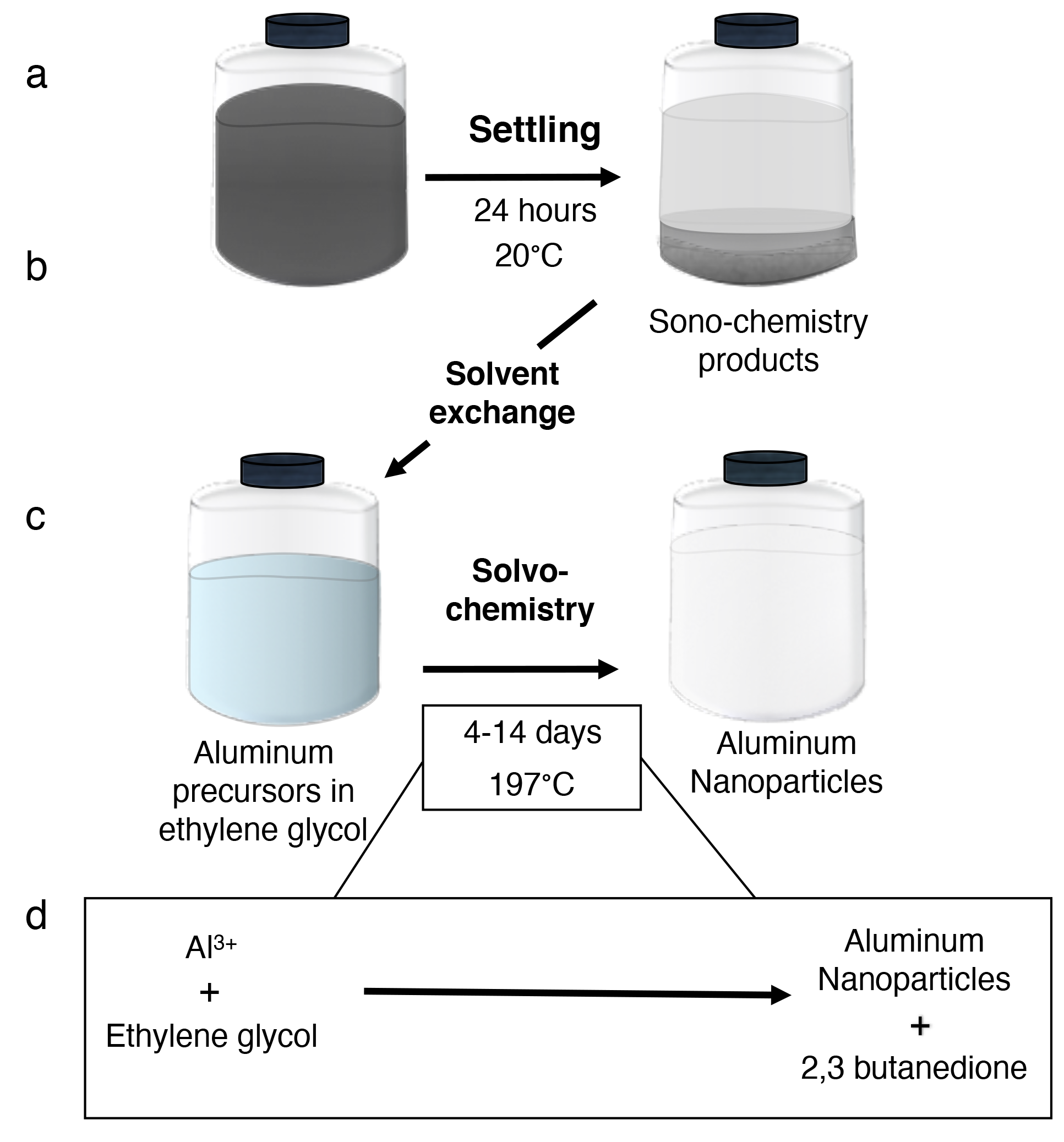}
\caption{(a) Settling step after sonochemistry to remove micro aluminum foils. (b) Solvent exchange, the residues of the sonochemistry step are replaced by clean ethylene glycol. (c)  solvochemistry step: the aluminum precursors react to form aluminum nanoparticles. (d) Streamline formula of the reaction.
}
\label{fig_chimie}
\end{figure}

The second step of the synthesis is solvochemistry (Figure \ref{fig_chimie}). It provides thermal energy to the system in order to reduce the aluminum precursors into nanoparticles, by precursors/solvents reactions and clusters aggregation. In their review, Lai \textit{et al.} state the different ways to obtain metallic nanocrystals by this method, changing the solvent, the precursor, the surfactant or the reductant and adding additives to obtain various shapes and sizes \cite{lai2015solvothermal}. Moreover, the duration and temperature of the reaction are two major factors. The flask containing the solution is left at room temperature for 24 hours. During this period, the micrometer-sized aluminum foils settle and can be removed by keeping only the supernatant (Figure \ref{fig_chimie}a). Then we proceed to a solvent exchange, using a homemade autoclave system under argon, to evaporate the solvent residues. A three-neck round bottom flask in a heating mantle is used as a reactor. One neck is connected to an air condenser, and linked to a Schlenk line under argon. The second neck is covered with a septum to allow for reactant injection. The third one is closed and not used. The gray supernatant of the settled solution is injected into the already purged flask (vacuum/argon cycles). The heating mantle is set at 110 °C to evaporate all volatile molecules from the solvent residues. These residues are eventually removed from the system by perforating the septum with a needle for 5 seconds every 2 minutes. We assume that all the volatile solvent are gone when the temperature of the flask reaches 110 °C. Then, we inject 60 mL of a pure anhydrous ethylene glycol (boiling point 197 °C) into the flask through the septum (Figure \ref{fig_chimie}b). From this point, the solvochemistry is performed during a maximum of 14 days at 197 °C under a slight argon over-pressure (Figure \ref{fig_chimie}c). The annealing time controls the diameter of the resulting Al nanoparticles, from 4 days for a 10 nm diameter to 14 days for 100 nm, as we will demonstrate later. As shown in Figure \ref{fig_chimie}d, during the heating period the aluminum precursors react with the oxidized form of ethylene glycol (ethanal, \ce{CH3-CHO}), following this sonochemistry reaction, inspired from \cite{yue2012ethylene}:

\ce{OH-CH2-CH2-OH ->[$197^\circ C$] CH3-CHO + H2O\\ 2 Al^{3+} + 6 CH3-CHO + 6 H2O -> 2 Al + 3 CH3-CO-CO-CH3 + 6 H3O+}

These reactions enable the creation of aluminum clusters which grow with the annealing time without emergence of new particles (up to 14 days). At the end of the reaction, we obtain aluminum nanoparticles and a residual 2,3 butanedione (\ce{CH3-CO-CO-CH3}), soluble in ethanol (Figure \ref{fig_chimie}c). The resulting solution is transparent with a yellowish color. Let us underline that the aluminum precursors created during the sonochemistry step could be replaced by commercial aluminum chloride precursors during the solvo-chemitry step. However in that case, we observed a synthesis of aluminum nanoparticles with a deteriorated quality (lower homogeneity). 

We emphasize that the oxidation of ethylene glycol creates water molecules. As the reaction occurs in a closed system, water accumulates in the flask with time. When the flask is saturated with water, the nanoparticle growth stops and ethylene glycol undergoes a caramelization reaction, forming carbon dots \cite{jaiswal2012cdots}. After 14 days of solvochemistry, we observed a saturation in the maximum size of the synthesized nanoparticles and an increase in the width of the size distribution. The solution also becomes fluorescent under blue-UV light excitation, indicating the presence of carbon dots. As a consequence, all syntheses were stopped at 14 days to avoid contamination of the solution with carbon dots.

\subsection{Purification}
In order to remove the byproducts created during the previous fabrication steps, a purification process is required, where the nanoparticles will be transferred in a new solvent. We chose absolute ethanol as a solvent for the Al nanoparticles to limit the oxidation of the nanoparticles in solution and allow a long time conservation. We perform two centrifugation steps in order to: (i) settle and remove the nano-foils \cite{shen2017purification} and (ii) allow for the solvent exchange. Owing to the low density of aluminum, we performed two successive centrifugation steps under an acceleration of 7800 RCF during 1 hour under a temperature set at 15 °C. First, 500 µL of the solution is mixed with 1 mL of absolute ethanol before the centrifugation cycle. The biggest particles settle down in the bottom of the flask and, owing to its larger density, ethylene glycol settles down simultaneously (at such concentrations, ethanol and ethylene glycol form an emulsion \cite{yue2012ethylene}). Then, we keep 500 µL of the supernatant and mix it with 500µL of absolute ethanol, before starting a second centrifugation cycle under the same conditions. After this last step, no large particles can be seen with the naked eye (the solution is clear), but we keep only 80\% of the supernatant. While the biggest particles and ethylene glycol settled down the tube and were removed, the smallest particles (the nanoparticles) are now diluted in pure ethanol. 

As the solvent is not anhydrous, a native oxide layer (alumina, \ce{Al2O3}) spontaneously forms at the surface of the nanoparticles as a shell. The hydroxylation of the alumina creates \ce{(Al)_n-OH} groups at the surface \cite{zhang2008structures}, with n ranging from 3 to 1. These groups can deprotonate and create negative charges at the surface of the nanoparticles. The Zeta potential in ethanol was measured at -29,8 mV. These surface charges prevent aggregation of the nanoparticles, yielding a long stability time of the solutions as demonstrated below.

\subsection{Size and morphological characterization}
\begin{figure}
\centering
\includegraphics[width=16cm]{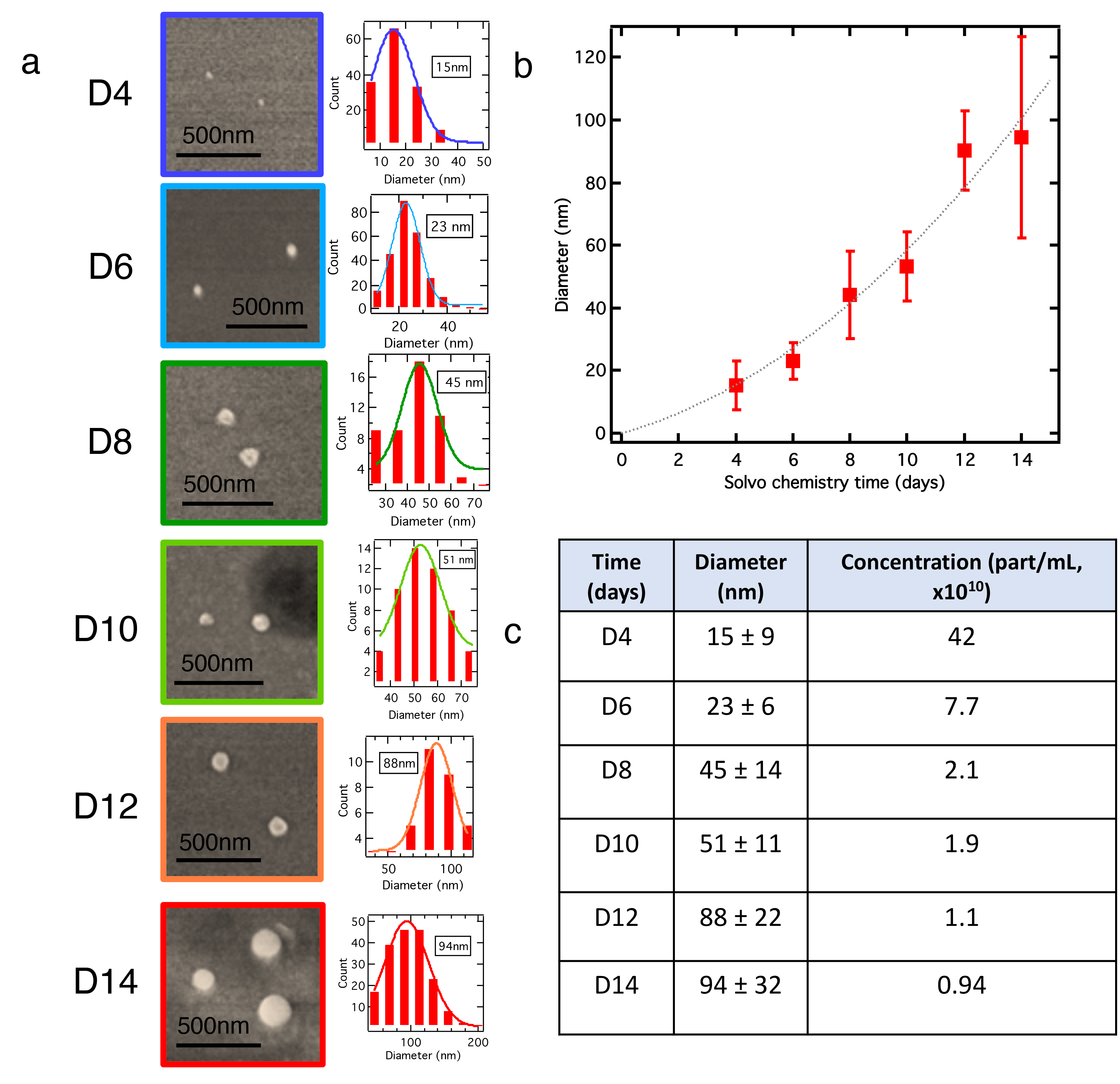}
\caption{(a) Left panel: SEM images of aluminum nanoparticles after 4 to 14 days of solvochemistry. Right panel: histograms of the measured diameters, fitted with a Gaussian function (solid line), for different reaction times. The average diameter (maximum of the Gaussian function) is shown as an inset. (b) Diameters of the aluminum nanoparticles depending on the solvochemistry time. (c) Table showing the calculated concentrations of the aluminum nanoparticles for each sample.}
\label{fig_taille}
\end{figure}

To rigorously characterize the growth of the Al nanoparticles, a regular sampling of the solution has been performed during the solvochemistry reaction. Each sample has been characterized with scanning electron microscopy (SEM). The observed aluminum nanoparticles exhibit a nice spherical shape (see Figure \ref{fig_taille}a). For each sample, between 50 and 200 particles have been systematically measured and their diameter are reported in the histograms of Figure \ref{fig_taille}a. Six samples, corresponding to the reaction times of 4, 6, 8, 10, 12 and 14 days, are presented. More SEM images with larger scale are presented in the supporting information (Fig. S3) for each reaction time. In each case, a SEM image and the corresponding diameter histogram are shown. In Figure \ref{fig_taille}b, we plot the diameter of the aluminum particles depending on the solvochemistry time, fitted by a polynomial curve in a dashed line as a guide for the eyes. The error bars are calculated from a Gaussian fit of the histograms and correspond to the full width at half maximum (FWMH). The size dispersion of the particles is reported in Figure \ref{fig_taille}c and is comparable with the results from Lu et al. \cite{lu2018polymer}, except for the sample taken after 14 days. For this sample, we observe the apparition of a new family of smaller particles as well as a larger size distribution. This observation corroborates the aforementioned process of caramelization, which is starting once the flask is saturated with the water produced during the reaction.

We also calculated the approximate concentration of nanoparticles in ethanol day by day using the Beer-Lambert law, as established by Ayala-Orozco \textit{et al.}, in the case of gold nanomatryoshkas \cite{ayala2014nanomatryoshkas}. For each sample, we acquired a UV-visible spectrum in solution (see Figure \ref{fig_stable}a) and the absorption value (Abs) was measured at the maximum of the resonance peak. To compute the concentration, we used the following formula:
\begin{equation}
    [\mathrm{part/mL}]=\dfrac{2.303 \times \mathrm{Abs}}{Q_{\mathrm{ext}} \, L} 
    \label{eq:part}
\end{equation}

In Equation \ref{eq:part}, Abs is the measured absorbance, and $L$ is the length of the optical path. The extinction cross-section $Q_{\mathrm{ext}}$ was calculated using Garcia de Abajo’s online Mie calculator \cite{de1999multiple} \cite{myroshnychenko2008modelling} which allows Mie calculations for core-shell (aluminum-alumina) geometries. For these calculations, we assumed a 3 nm alumina shell (as measured by transmission electron microscopy (TEM), see below) and used the average diameters shown in Figure \ref{fig_taille}a. The calculated concentrations are around $10^{10}$ part/mL (see Figure \ref{fig_taille}c for details). These concentrations are comparable with the results obtained by Jana \textit{et al.} for the synthesis of gold nanoparticles \cite{jana2001} and by Renard \textit{et al.} for aluminum nanocrystals synthesis \cite{renard2020uv}. Note that the concentration decreases during the growing process, probably due to the adsorption of aluminum clusters and nanoparticles onto the flask’s glass wall. With this method, we are limited to 90 nm diameter while keeping a good dispersion. Following the dashed line in Figure \ref{fig_taille}b, we can expect that particles smaller than 10 nm can be synthesized with shorter solvochemistry time, but we were not able to observe them with SEM. 

\subsection{Structural characterization} 
\begin{figure}
\centering
\includegraphics[width=16cm]{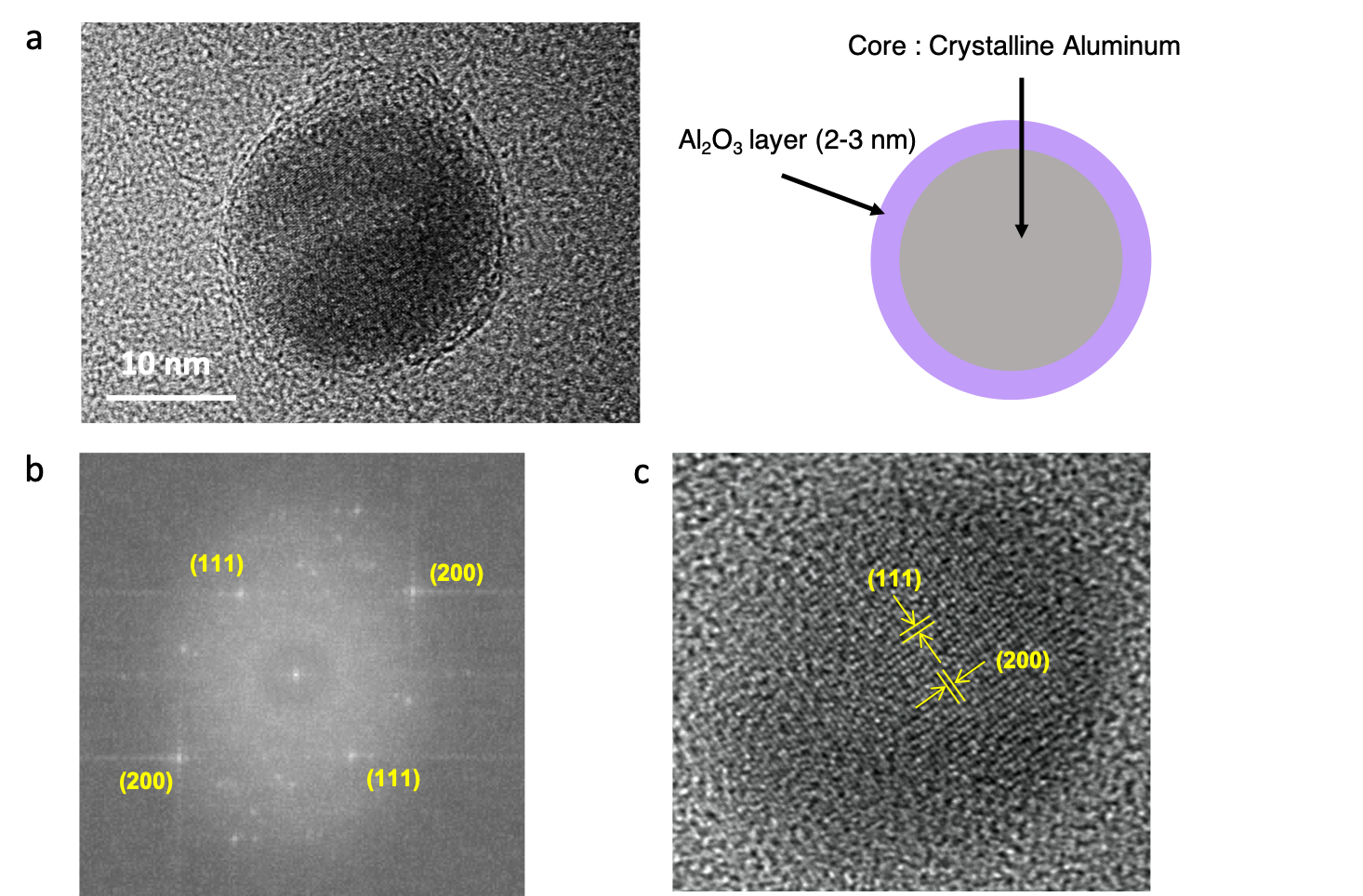}
\caption{(a) TEM image of a 20 nm diameter aluminum nanoparticle (dark) with the alumina shell (in light grey) and the corresponding schema. (b) Diffraction pattern. (c) High-resolution TEM image showing the (111) and (200) atomic planes of aluminum.}
\label{fig_tem}
\end{figure}

Aluminum has a low oxidation resistance. This naturally creates an alumina passivation layer, avoiding bulk oxidation of the nanoparticles. We performed high-resolution TEM imaging in order to measure the thickness of this oxide layer. Figure \ref{fig_tem}a shows a TEM image of a 20 nm diameter nanoparticle. This particle has been stabilized more than two months before the measurements. In Figure \ref{fig_tem}a, we can easily observe a 2-3 nm thick oxide layer at the surface of the particle. This small thickness is enough to avoid further oxidation. Figure \ref{fig_tem}c shows the highly crystalline aluminum core with two visible atomic planes (111) and (200), also visible on the diffraction patterns shown in Figure \ref{fig_tem}b. The nanoparticles are then core-shell structures, with crystalline aluminum core and alumina shell. We can assume that the 2-3 nm thickness is present on all the samples, and that the observed diameter in Figure \ref{fig_taille}a includes the oxide shell, an assumption supported by many studies in literature describing the formation and the stabilization of the alumina shell on aluminum nanoparticles, either in solution and air \cite{rai2006understanding,knight2014aluminum,langhammer2008localized,zhang2019long}.
Calculations of concentration in Figure \ref{fig_taille}c and Mie simulations in Figure \ref{fig_optic} both take this thickness into account. 

\subsection{Optical characterization} 
\begin{figure}
\centering
\includegraphics[width=8cm]{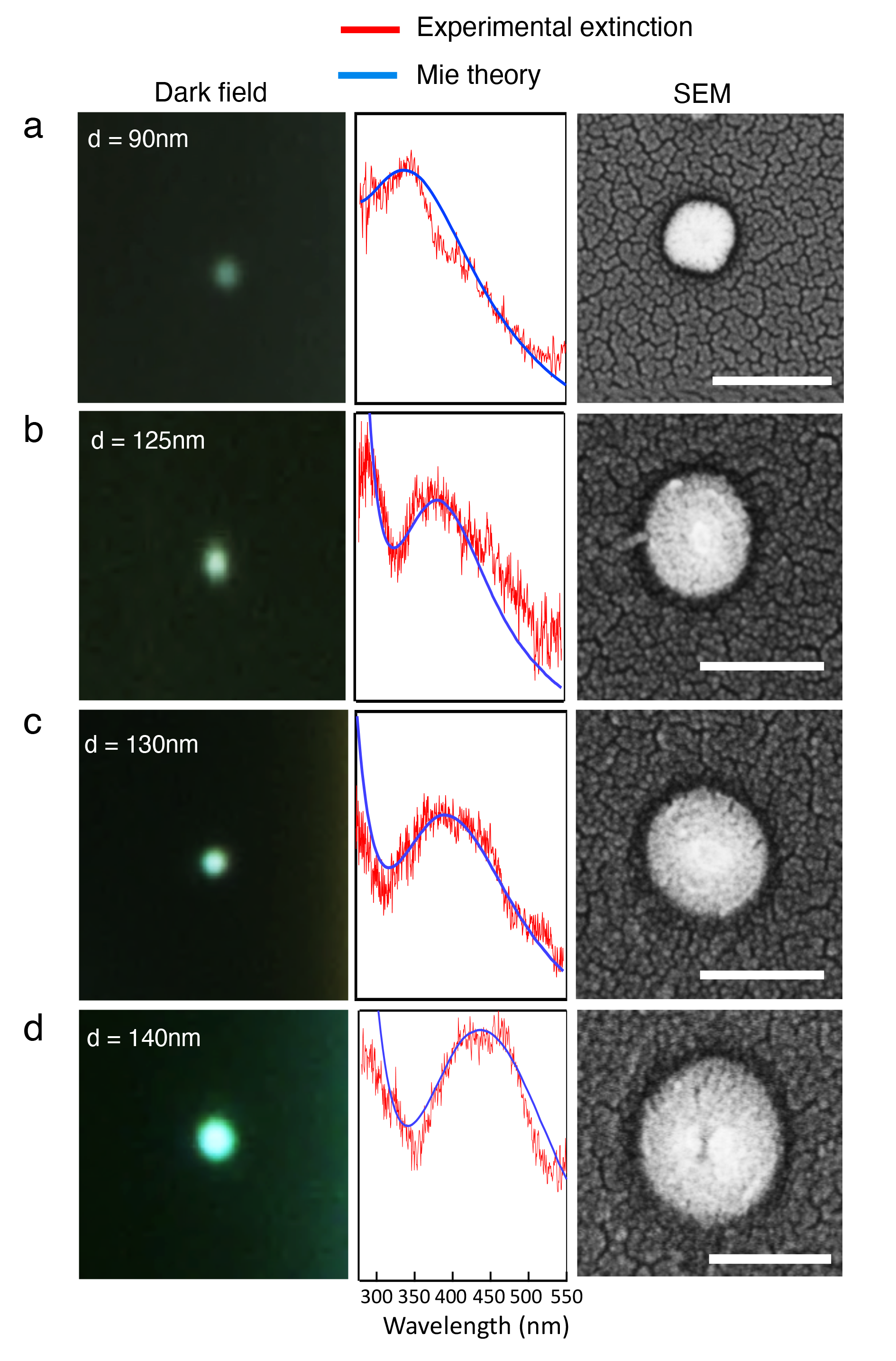}
\caption{Optical characterization of several Al nanoparticles obtained after 14 days of solvochemistry. Left column: dark field image of single aluminum nanoparticles. Middle column: raw normalized extinction spectra on a single nanoparticle (red line) and corresponding Mie calculation (blue line). Right column: corresponding SEM images. Each line of the figure shows the same particle in every case, with a diameter of (A) 90 nm, (B) 125 nm, (C) 130 nm and (D) 140 nm.
}
\label{fig_optic}
\end{figure}

To corroborate the structural and topography information performed on our aluminum nanoparticles, we study their plasmonic responses at the single nanoparticle scale using a homemade optical setup. To decrease the surface density of Al nanoparticles on the sample, the nanoparticle-containing solution was strongly diluted in ethanol before being deposited on a fused-silica microscope slide. A grid was imprinted on the slide beforehand, allowing an easier localization of the particles. Hence, the grid allows us to correlate the optical and SEM imaging, ensuring we are looking at the same particle in each case. Dark field images of four nanoparticles with different diameters (90, 125, 130 and 140 nm) are presented on the left column in Figure \ref{fig_optic}. All these particles were obtained after 14 days of solvochemistry, but different diameters are observed due to size dispersion (see the histogram in Figure \ref{fig_taille}a, bottom row). We observe a change in the scattered color with the diameter of the nanoparticle (from dark green to turquoise), reflecting the interaction between the light and those particles.

Extinction spectroscopy was performed on the same particles, using the UV homemade transmission extinction setup. The particles were deposited onto a transparent substrate and illuminated with a high-stability, broad spectrum white lamp (Energetiq Laser-Driven Light Source, Hamamatsu). Incident light is slightly focused on the sample by a 5x, NA=0.12 UV objective and the transmitted light is collected through another x80, NA=0.5 UV objective. Confocal spatial filtering is performed using the 200 µm diameter core of an UV optical fiber, which is connected to a spectrometer equipped with an UV grating (Fergie, Princeton Instruments). This results in a collection area of 2.5 µm diameter, making it possible to obtain the extinction spectra of \textit{single} nanoparticles provided they are dispersed enough on the substrate. Taking into account the transmission of all optics, the operating spectral range of the setup is 250 - 600 nm. This setup configuration is currently the best compromise (beam-splitter, objective, fiber, source, spectrometer) between a broad wavelength range and a signal-to-noise ratio high enough to measure a single aluminum nanoparticle. Raw single nanoparticle extinction spectra are shown in red on the second column in Figure \ref{fig_optic}, for the four different diameters. The complexity of the measurement resides on the weak extinction cross-section of aluminum. However, we are able to show the redshift of the local surface plasmonic resonance with the increase of the diameter. In some raw spectra, a shoulder can be observed next to the main resonance. These shoulders appear to depend on the experimental conditions, especially on the focus, and cannot be attributed to a plasmonic resonance. We attribute them to a slight change of the objective-sample distance (defocus) between the measurements of the reference and of the signal of the single nanoparticle. Moreover, we validate these experimental spectra by overlapping the theoretical spectrum (blue curves in Figure \ref{fig_optic}), as obtained by Mie calculation on an alumina-coated aluminum nanosphere using the Palik dispersion curves \cite{de1999multiple} \cite{myroshnychenko2008modelling}. A very good agreement between the raw data and the Mie theory is observed. This further confirms the good sphericity of the nanoparticles. Then, each particle was imaged with SEM after a 4 nm gold/palladium layer metallization. Each SEM image gives us the real diameter of the obtained nanoparticles. These measured diameters were used in the previous Mie calculations. 

\subsection{Stability} 
\begin{figure}
\centering
\includegraphics[width=12cm]{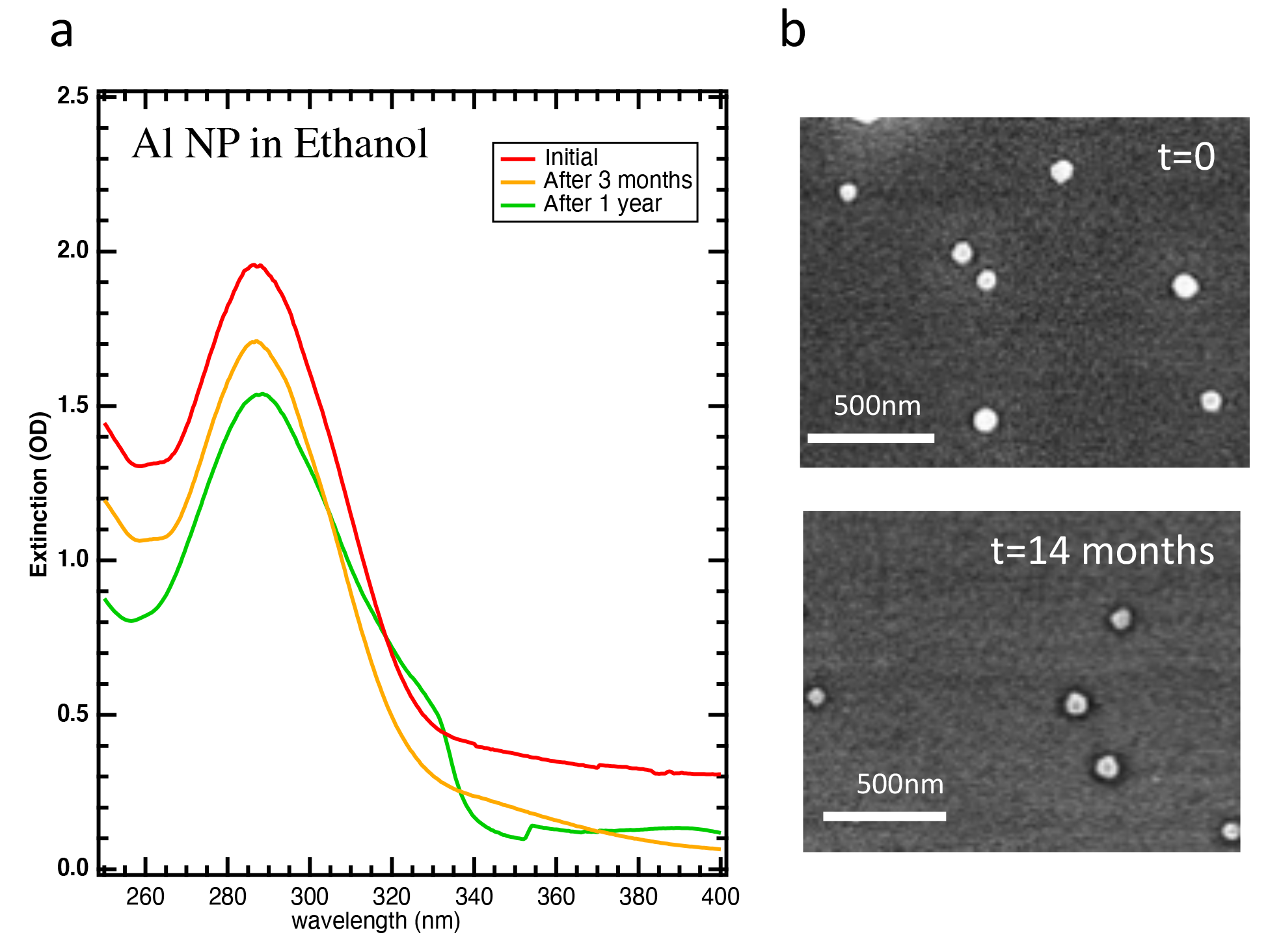}
\caption{(a) Extinction spectra of the same aluminum nanoparticles solution in ethanol at different stabilisation time: initial, 3 months and 1 year. (b) SEM observation of the initial solution (top image) and the same solution after 14 months (bottom image).}
\label{fig_stable}
\end{figure}

One of the drawbacks of the metallic colloidal synthesis is the poor stability in solution. For example, the stability of commercial gold nanoparticles is usually guaranteed for two months only. To avoid the aggregation phenomena and increase the stability in time, we usually functionalize the surface with ligands \cite{kanninen2008copper}, polymers \cite{manson2011polyethylene} or surfactant molecules \cite{pisarvcik2018silver}. By controlling the pH of the solution, it is possible to maintain the particles' optical properties for months. In our synthesis, the surface of the aluminum nanoparticles is protected by a negatively charged aluminum oxide. This charge avoids the aggregation process and we can keep our nanoparticles in ethanol for months. We can see in Figure \ref{fig_stable}a that the plasmon resonance after 3 months is equivalent to the original, with a lower absorbance. Without modification of the spectra, we can assume that nanoparticles are stuck to the glass of the container, but no aggregations are visible. But after one year, we can see a beginning of agglomeration in the spectra, with a bump at 330 nm. In Figure \ref{fig_stable}b, we can see on SEM images that, even after 14 months, the nanoparticles shape and size are still the same than the original solution in majority, and the solution is still usable. 

\section{Conclusion}
In conclusion, we demonstrated a novel elaboration path for the production of spherical aluminum nanoparticles, with diameters ranging from 10 nm to 100 nm. The synthesis involves two steps: first sonochemistry to create the aluminum precursors, and then solvochemistry to reduce the precursors and create nanoparticles. The solvochemistry reaction time controls the diameter of the particles. We characterized the spherical shape, the concentration and the diameter dispersion. The morphological characterization shows that the particles are composed of a highly crystalline metallic core with a thin oxide shell. This native shell is an advantage for the nanoparticles' stability in solution: we demonstrated that the particles are stable for 1 year in ethanol, with negligible aggregation. Finally, we showed the plasmonic optical response of single aluminum nanoparticles. Correlations were made between dark field imaging, extinction spectroscopy and SEM imaging on four different particles, and were corroborated by Mie theory for a core-shell nanoparticle with 3 nm alumina shell. We emphasize that this elaboration path is compatible with mass production, as up-scaling the production would only requires a larger ultra-sonic bath. 

Albeit we demonstrated here only the synthesis of spherical nanoparticles, we underline that other shapes are between reach, as truncated octahedron nanoparticles which ascertain the cristallinity of the synthesized objects. For instance, the addition of trioctylphosphine oxide (TOPO) during the solvochemistry should allow to control the NP growth into specific directions \cite{hou2005size}.


\begin{acknowledgement}
We are most grateful to Francesco Bisio for his contribution to the characterization of the nanoparticles.
 
\end{acknowledgement}

\section*{Funding Sources}
Financial support of Nano’Mat (www.nanomat.eu) by the "Ministère de l'enseignement supérieur et de la recherche", the “Fonds Européen de Développement Régional" (FEDER), the "Région Grand-Est", and the “Conseil général de l’Aube” are acknowledged.

This project was supported by the Agence nationale de la recherche (ANR), under project SMFLUONA (grant ANR-17-CE11-0036). This work has been supported by the EIPHI Graduate School (grant ANR-17-EURE-0002). This work has been made within the framework of the Graduate School NANO-PHOT (grant ANR-18-EURE-0013).

\begin{suppinfo}
Short movie clearly explaining all of the procedures to produce our aluminum nanoparticles (AVI).
A supporting information is also available (PDF). This material is available free of charge via the internet at http://pubs.acs.org. The supporting information contains three parts: S1 Infrared spectroscopy spectra of the solvents after the sonochemistry process. S2 Conductivity of the solution during and after the sonochemistry process. S3 SEM images of the nanoparticles at different sonochemistry process time. 
\end{suppinfo}

\bibliography{NP-synthesis}

\end{document}